\documentstyle[aps,prl,epsfig]{revtex}
\begin{document}
\newcommand{\be}{\begin{equation}}
\newcommand{\ee}{\end{equation}}
\newcommand{\bel}[1]{\begin{equation}\label{#1}}
\newcommand{\bea}{\begin{eqnarray}}
\newcommand{\eea}{\end{eqnarray}}
\newcommand{\ba}{\begin{array}}
\newcommand{\ea}{\end{array}}
\newcommand{\bra}[1]{\mbox{$\langle \, {#1}\, |$}}
\newcommand{\ket}[1]{\mbox{$| \, {#1}\, \rangle$}}
\newcommand{\exval}[1]{\mbox{$\langle \, {#1}\, \rangle$}}
\tighten
%%%%%%%%%%%%%%%%%%%%%%%%%%%%%%%
%\onecolumn
%\twocolumn[\hsize\textwidth\columnwidth\hsize\csname @twocolumnfalse\endcsname

\title{Steady-state selection in driven diffusive systems with open boundaries}
\author{ Vladislav Popkov$^{1,2}$  and Gunter M. Sch\"utz$^1$ }
\address{$^1$Institut f\"ur Festk\"orperforschung, Forschungszentrum J\"ulich,
52425 J\"ulich, Germany\\
$^2$Institute for Low Temperature Physics,310164 Kharkov,Ukraine}
\maketitle
\begin{abstract}
We investigate the  stationary states of one-dimensional driven
diffusive systems, coupled to boundary reservoirs with fixed particle 
densities. We argue that the generic phase diagram is governed by an
extremal principle for the macroscopic current irrespective of the
 local dynamics. In  particular, we predict a
 minimal current phase for  systems with local minimum  
in the current--density relation. This phase is explained by
a dynamical phenomenon, 
the  branching and coalescence of shocks, Monte-Carlo simulations confirm
the theoretical scenario.
\end{abstract}
\pacs{PACS: 05.70.Fh, 05.70.Ln, 02.50.Ga}
% 05.70.Ln  Nonequilibrium thermodynamics, irreversible processes
% 05.70.Fh  Phase transitions: general aspects
]

%\newpage

A recurrent problem in the investigation of many-body systems far from
equilibrium is posed by the coupling of a driven particle system with
locally conserved particle number to external reservoirs with which
the system can exchange particles at its boundaries. In the presence of
a driving force a particle current will be maintained and hence the system
will always remain in an nonequilibrium stationary state characterized by
some bulk density and the corresponding particle current. While for
periodic boundaries the density is a fixed quantity, the experimentally
more relevant scenario of open boundaries
naturally leads to the question of steady-state selection, i.e. the question
which stationary bulk density the system will assume as a function of the
boundary densities \cite{N1}. In the topologically simplest case of quasi
one-dimensional systems this is of
importance for the understanding of many-body systems in which the dynamic
degrees of freedom reduce to effectively one dimension as e.g. in traffic flow
\cite{traffic}, kinetics of protein synthesis \cite{protein},
or diffusion in narrow channels \cite{Kukl96}.

Within a phenomenological approach this problem was first addressed in general
terms by Krug \cite{Krug91} who postulated a maximal-current principle
for the specific case where the density $\rho^+$ at the right boundary to
which particles are driven is kept at zero. The exact solution of the totally
asymmetric simple exclusion process (TASEP) \cite{TASEP} for arbitrary
left and right boundary
densities $\rho^-,\rho^+$ confirms and extends the results by Krug. The
complete phase diagram comprises boundary-induced phase transitions of first
order between a low-density phase and a high density phase, and a
second-order transition from these phases to a maximal current phase
\cite{Ligg75,Schu93b,Derr}. Analysis of the density profile \cite{Schu93b}
provides insight into the dynamical mechanisms that lead to these phase
transitions and shows that the phase diagram is generic
for systems with a single maximum in the current-density relation
\cite{Kolo98}. Experimental evidence for the first-order transition is found
in the process of biopolymerization for which the TASEP with open boundaries
was originally invented as a simple model \cite{protein2}, and more directly
also in recent measurements of highway traffic close to on-ramps
\cite{Neub99,Popk99}. Renormalization group studies \cite{Oerd98} indicate
universality of the second order phase transition.

In this Letter we develop a dynamical approach to generic driven one-component
systems with {\em several} maxima in the current-density relation, and we show
that a novel phase of rather unexpected nature appears: For a certain
range of boundary densities the steady state carries the
minimal current between two maxima even though both boundary
densities support a higher current. We shall refer to this phase as minimal
current phase. More generally, we claim that the current always obeys the
extremal principle
\bea
\label{max}
 j & = &
\max_{\rho \in [\rho^{+},\rho^{-}]} j(\rho) \mbox{ for } \rho^{-}>\rho^{+} \\
\label{min}
 j & = &
\min_{\rho \in [\rho^{-},\rho^{+}]} j(\rho) \mbox{ for } \rho^{-}<\rho^{+}.
\eea

To understand the origin of this extremal principle we first note that in the
absence of detailed balance stationary behavior cannot be understood in terms
of a free energy, but has to be derived from the system dynamics. For
definiteness consider a driven lattice gas with hard-core repulsion. At
$\rho=1$ no hopping can take place and hence the current vanishes. Two local
maxima can arise as the result of sufficiently strong repulsion between
nearest neighbor particles as opposed to the pure on-site repulsion of the
usual TASEP which leads to a single maximum (Fig. \ref{curr}).

At first sight one might not expect such a little change in the interaction
radius of the particles to affect the phase diagram. However, the extremal
principle (\ref{max}), (\ref{min}) predicts that
the full phase diagram (Fig.~\ref{phase}) generically
consists of seven distinct phases, including two maximal current phases with
bulk densities corresponding to the respective maxima of the current
\cite{maxcoex} and the minimal current phase in a regime defined by
\be
j(\rho^{+}),j(\rho^{-}) > j(\rho_{min}); \ \
\rho^{-} < \rho_{min} < \rho^{+}
\label{min_condition}.
\ee
Here the system organizes
itself into a state with bulk density $\rho_{bulk}$ corresponding to the local
minimum of the current. As in the maximal current phases no fine-tuning of the
boundary densities is required.

In order to understand the basic mechanisms which determine the steady state
selection we first follow Ref. \cite{Kolo98} and consider the collective
velocity
\bel{1}
v_c = j'(\rho)
\ee
which is the velocity of the center of mass of a local perturbation
in a homogeneous, stationary background of density $\rho$ (Fig. 3(a)).
A second quantity of interest is the shock velocity
\bel{2}
v_s = \frac{j_2 - j_1}{\rho_2 - \rho_1}
\ee
of a `domain wall' between two stationary regions of densities $\rho_{1,2}$
(Fig. 3(b)). Notice that both velocities may change sign, depending on the
densities.\\

These velocities are sufficient to understand the phase diagram of systems
with a single maximum in the current \cite{Kolo98}. Further to these notions
we need to introduce here the concepts of coalescence and branching of shocks.
A single large shock (with a large density difference $\rho_2-\rho_1$) may be
understood as being composed of subsequent smaller shocks with narrow plateaux
at each level of density (Fig.~\ref{schema}). In the usual asymmetric exclusion
process these shocks travel with different relative speeds such that they
coalesce and form a `bound state' equivalent to a single shock
\cite{Ferr98}. In the present situation, however, the minimum in the
current-density relation leads to more complicated dynamics. A closer
investigation of Eq. (\ref{2}) shows that depending on
$\rho_{1,2}$ a single shock may branch into two distinct shocks,
moving away from each other, discussed below.

With these observations the dynamical origin of the phase transition lines
can be understood by considering the time evolution of judiciously chosen
initial states. Because of ergodicity, the steady state does not
depend on the initial conditions and a specific choice involves no loss of
generality. We turn our attention to a line $\rho^+ = c$ with $\rho_{min} < c
<\rho_2^{\ast}$ in the phase diagram which crosses the minimal current phase.
Along this line it is convenient to consider an initial configuration
with a shock with densities $\rho^-$ and $\rho^+$ on the left and on the right
respectively, which is composed of many narrow subsequent shocks
at various levels of intermediate densities (Fig.~\ref{schema}).

(i) Let us start with the point with equal boundary densities
in which case the system evolves into a steady state with the same
bulk density $\rho_{bulk}=\rho^-=\rho^+$.

(ii) Lower $\rho^-$ slightly below $\rho^+$ with just a single shock separating
both regions. According to (\ref{2}) the shock travels with speed $v_s=
(j^{+}-j^{-})/(\rho^{+}- \rho^{-}) \ >0$ to the right, making the bulk density
equal to $\rho^{-}$. At the same time, small disturbances will according
to (\ref{1}) also drift to the right, as $v_c= j'(\rho^-) >0$ in
this region, thus stabilizing the single shock.

(iii) Now, lower $\rho^-$ below $\rho_{min}$. While the shock
velocity $v_s$ is still positive, so that one expects the shock
to move to the right, the collective velocity  $v_c= j'(\rho^-) <0$
indicates that disturbances will spread to the left. This discrepancy
marks the failure of a single shock scenario. In order to resolve
it, consider for simplicity  $\rho^-=  \rho_1^*$ 
and return to the picture with many subsequent shocks at each density
level between $\rho^-$ and $\rho^+$ (Fig.\ref{schema}). Eq. (\ref{2})
shows that  small shocks below $\rho_{min}$ will move to the left, while
 those above $\rho_{min}$ will move to the 
right\footnote{Actually, the small shock between the levels 
$\rho_k^-,\rho_k^+$ will be stable only if
$v_c(\rho_k^+)< v_s < v_c(\rho_k^-) $, see Eq.(\ref{1},\ref{2}). 
Some of small shocks satisfy the above condition and some do not.
This brings about additional structure to the resulting shock,
which will be discussed elsewhere
 }. 
The leftmost of the left-moving shocks
will merge in a single one, and so will the rightmost of
the right-moving shocks.
The result are two single shocks and
 moving different directions and thus expanding
the region with the density $ \rho_{bulk}=\rho_{min}$. The system
is in the {\it minimal current phase}.
 This picture is well supported by the Monte-Carlo simulations
shown in Fig.~\ref{shock}, demonstrating the branching of a single shock
into two distinct shocks moving in opposite directions.
The  minimal current phase will persist for any left boundary density
in the range  $\rho^- \in [\tilde \rho_1,\rho_{min}]$.
Notice that the change of bulk density is continuous across the point
$\rho^- = \rho_{min}$ to the minimal current phase, so the
transition is of the second order.

(iv) As we lower $\rho^-$ below $\tilde \rho_1$, the shock
velocity $v_s =(j_{min}-j(\rho^-))/(\rho_{min}-\rho^-) >0$ becomes positive.
The shock is moving to the right, leading to a low density phase with bulk
density $ \rho_{bulk}=\rho^-$ which drops discontinuously from
$\rho_{bulk}=\rho_{min}$ at $\rho^-=\tilde \rho_1+0$ to $\rho_{bulk}=\rho^-$
at $\rho^-=\tilde \rho_1-0$. The system undergoes a first order phase
transition. On the transition line the shock performs an unbiased random
walk, separating coexisting regions of densities $\rho_{min}$ and
$\rho^-$ respectively.
% marked by a bold line in the phase diagram Fig. \ref{phase}.
%Note that for all the  range of $\rho^- \in [0,\rho^+]$
% considered so far, the steady bulk density is the one corresponding
%to the minimum of current in the region $[\rho^-,\rho^+])$.

(v) Let us start again from $\rho^-=\rho^+$ and now increase $\rho^-$.
Until one reaches $\rho^- =  \rho^*_2$, the collective velocity
$v_c= j'(\rho^-) >0$ is positive, leading to $\rho_{bulk}=\rho^-$.

(vi) As soon as $\rho^{-}$ crosses the point  $\rho^{-} = \rho^*_2$, the sign
of collective velocity  $v_c$ changes
and the overfeeding effect \cite{Schu93b} occurs: a perturbation from the left
does
not spread into the bulk \cite{Kolo98} and therefore further increase of the
left boundary density does not increase the bulk density. The system enters
the maximal current phase II through a second-order transition.

Using analogous arguments one constructs the complete phase diagram
(Fig.~\ref{phase}) and obtains the extremal principle (\ref{max},
\ref{min}). The velocities (\ref{1}), (\ref{2}) which determine the
phase transition lines follow from the current-density relation.
This behavior can be checked with Monte Carlo simulations.
A model with two maxima of the current is a TASEP with next nearest
neighbor interaction defined by the bulk hopping rates
(see A26,A27 in \cite{Katz84})
\bea
0\;1\;0\;0 & \to & 0\;0\;1\;0\hspace{1cm}
\mbox{ with rate } 1+\delta \label{rate1}\\
1\;1\;0\;0 & \to & 1\;0\;1\;0\hspace{1cm}
\mbox{ with rate } 1+\epsilon \label{rate2}\\
0\;1\;0\;1 & \to & 0\;0\;1\;1\hspace{1cm}
\mbox{ with rate } 1-\epsilon \label{rate3}\\
1\;1\;0\;1 & \to & 1\;0\;1\;1\hspace{1cm}
\mbox{ with rate } 1-\delta \label{rate4}
\eea
with $|\epsilon| <1; \ \ |\delta|<1$.
For sufficiently strong repulsive interaction $1-\epsilon \ll 1$
the current at half-filling is strongly suppressed, which brings
about two maxima structure in the
current-density relation (Fig.~\ref{curr}). The limit 
$\epsilon=1$ leads to $j_{min}=0$.  For negative $\epsilon$ 
(attractive interaction) the current-density relation always has 
a single maximum. The other parameter, $\delta$,
is responsible for the particle-hole symmetry. $\delta=0$ corresponds
to symmetrical graph $j(\rho)=j(1-\rho)$.  $\delta \neq 0$ breaks
the particle-hole symmetry in favor of larger particles current
( $\delta > 0$) or larger vacancies current ( $\delta < 0$).
In particular,  $\delta \neq 0$ is responsible for different heights
of two maxima on Fig.~\ref{curr}.

The injection at the left boundary site 1 and extraction of particles
at the right boundary site $L$ is chosen to correspond to coupling to boundary
reservoirs with densities $\rho^\pm$ respectively. Along the line
$\rho^+=\rho^-$ the stationary distribution is then exactly given by the
equilibrium distribution of an 1-D Ising model with boundary fields and the
bulk field such that the density profile is constant with density
$\rho=\rho^+=\rho^-$ \cite{Anta99}. The current
$j = (1+\delta) \exval{0100} + (1+\epsilon) \exval{1100} +
(1-\epsilon) \exval{0101} + (1-\delta) \exval{1101}$
as a function of the density can be calculated exactly
using standard transfer matrix techniques. The exact graph is shown in
Fig.~\ref{curr} for specific values of the hopping rates.

We performed Monte Carlo simulations for systems of sizes
$L$ from 100 to 1000. Densities and currents were averaged over at least
 50$L$ rounds, and averaged over 7 different histories. We
consider the bulk density $\rho_{bulk}$ as order parameter. 
In analogy to equilibrium phase transitions a singularity (jump
discontinuity) in the first derivatives of $\rho_{bulk}$ signals 
a first (second) order phase transition. In the finite system
we determined the location of the first (second) order transition
by the appearance of peak (jump) in the first derivatives
of the bulk density $\rho_{bulk} (\rho^{+}, \rho^{-})$ with
respect to  $\rho^{+}$ and $\rho^{-}$. As initial state
we chose either the empty or the completely filled lattice, whichever
gave the faster convervence. Despite finite size effects,  precise
analysis of which requires further investigations,
the overall agreement of the simulated phase diagram
with the predicted one is very good already for $L=150$ 
(Fig.\ref{phase})\footnote{ We have
chosen the set of parameters $\epsilon$ and $\delta$, which
shows distinct minimum between two maxima. For the other
choices of $\epsilon, \delta$, the phase diagram can be obtained in
the same way, e.g. by current-density relation and extremal 
principles Eqs.(\ref{max},\ref{min}).}.

We conclude that the interplay of density fluctuations and shock diffusion,
coalescence and branching resp. as described above represents the basic
mechanisms which determine the
steady state selection (\ref{max},\ref{min})
of driven diffusive systems with a nonlinear current--density relation. 
A surprising phenomenon is the occurrence of
the self-organized minimal current phase. Since little reference is made
to the precise nature of the dynamics we argue that the phase diagram is
generic and hence knowledge of the macroscopic current-density
relation of a given physical system is sufficient to calculate the exact
nonequilibrium phase transition lines which determine the density of the
bulk stationary state. For systems with more than two maxima in the current
the interplay of more than two shocks has to be considered.

A fundamentally different scenario seems likely in systems with long-range
bulk correlations, which, however, is untypical in one dimension. Perhaps the
most interesting open question which one might address in a similar fashion
is the question of steady-state selection in systems with two or more local
conservation laws. This could provide insight into the mechanisms responsible
for boundary-induced spontaneous symmetry breaking \cite{sym} in 1-D driven
diffusive systems with open boundaries.

{\bf Acknowledgements} This work was partially supported 
by DAAD and CAPES within the PROBRAL programme. G.M.S. thanks 
H. Guiol for  useful discussions.
% during his stay in  IME/USP. 

\newpage

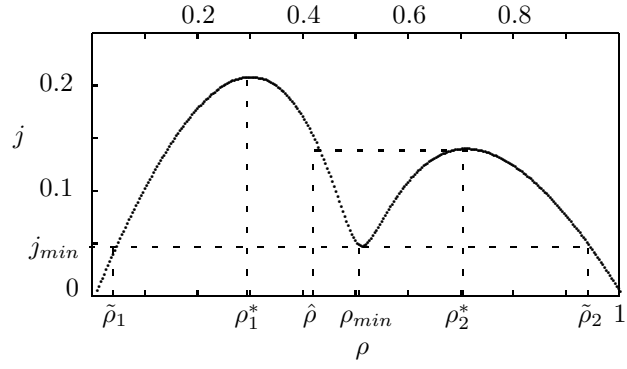
\begin{figure}
\setlength{\unitlength}{0.7mm}
\begin{center}
%\begin{picture}(100,60)(0,-10)
\begin{picture}(100,60)(0,-10)

\put(0,0){\line(0,1){50}}
\put(100,0){\line(0,1){50}}
\put(0,0){\line(1,0){100}}
\put(0,50){\line(1,0){100}}
\put(-15,29){$j$}
\put(50,-11){$\rho$}

\multiput(10,0)(10,0){9}{\line(0,1){1}}
\multiput(10,50)(10,0){9}{\line(0,-1){1}}

\put (-5,0){$0$}
\put (-10,19){$0.1$}
\put (-10,39){$0.2$}

\put(-12,8){$j_{min}$}
\put (17,52){$0.2$}
\put (37,52){$0.4$}
\put (57,52){$0.6$}
\put (77,52){$0.8$}
\put (99,-5){$1$}

\put(0,10){\line(1,0){1}}
\put(0,20){\line(1,0){1}}
\put(0,30){\line(1,0){1}}
\put(0,40){\line(1,0){1}}

\put(0.333333,0.797333){.}
\put(0.666667,1.589297){.}
\put(1.000000,2.375838){.}
\put(1.333333,3.156901){.}
\put(1.666667,3.932431){.}
\put(2.000000,4.702371){.}
\put(2.333333,5.466664){.}
\put(2.666667,6.225254){.}
\put(3.000000,6.978080){.}
\put(3.333333,7.725084){.}
\put(3.666667,8.466206){.}
\put(4.000000,9.201385){.}
\put(4.333333,9.930559){.}
\put(4.666667,10.653665){.}
\put(5.000000,11.370639){.}
\put(5.333333,12.081418){.}
\put(5.666667,12.785936){.}
\put(6.000000,13.484127){.}
\put(6.333333,14.175923){.}
\put(6.666667,14.861257){.}
\put(7.000000,15.540059){.}
\put(7.333333,16.212259){.}
\put(7.666667,16.877787){.}
\put(8.000000,17.536570){.}
\put(8.333333,18.188535){.}
\put(8.666667,18.833608){.}
\put(9.000000,19.471713){.}
\put(9.333333,20.102775){.}
\put(9.666667,20.726716){.}
\put(10.000000,21.343457){.}
\put(10.333333,21.952919){.}
\put(10.666667,22.555021){.}
\put(11.000000,23.149680){.}
\put(11.333333,23.736814){.}
\put(11.666667,24.316338){.}
\put(12.000000,24.888166){.}
\put(12.333333,25.452211){.}
\put(12.666667,26.008384){.}
\put(13.000000,26.556597){.}
\put(13.333333,27.096758){.}
\put(13.666667,27.628774){.}
\put(14.000000,28.152552){.}
\put(14.333333,28.667997){.}
\put(14.666667,29.175012){.}
\put(15.000000,29.673498){.}
\put(15.333333,30.163356){.}
\put(15.666667,30.644485){.}
\put(16.000000,31.116782){.}
\put(16.333333,31.580143){.}
\put(16.666667,32.034461){.}
\put(17.000000,32.479629){.}
\put(17.333333,32.915538){.}
\put(17.666667,33.342077){.}
\put(18.000000,33.759132){.}
\put(18.333333,34.166589){.}
\put(18.666667,34.564333){.}
\put(19.000000,34.952244){.}
\put(19.333333,35.330203){.}
\put(19.666667,35.698087){.}
\put(20.000000,36.055773){.}
\put(20.333333,36.403134){.}
\put(20.666667,36.740043){.}
\put(21.000000,37.066369){.}
\put(21.333333,37.381981){.}
\put(21.666667,37.686743){.}
\put(22.000000,37.980521){.}
\put(22.333333,38.263174){.}
\put(22.666667,38.534562){.}
\put(23.000000,38.794542){.}
\put(23.333333,39.042968){.}
\put(23.666667,39.279692){.}
\put(24.000000,39.504564){.}
\put(24.333333,39.717431){.}
\put(24.666667,39.918138){.}
\put(25.000000,40.106526){.}
\put(25.333333,40.282436){.}
\put(25.666667,40.445703){.}
\put(26.000000,40.596163){.}
\put(26.333333,40.733646){.}
\put(26.666667,40.857982){.}
\put(27.000000,40.968996){.}
\put(27.333333,41.066512){.}
\put(27.666667,41.150349){.}
\put(28.000000,41.220325){.}
\put(28.333333,41.276253){.}
\put(28.666667,41.317946){.}
\put(29.000000,41.345212){.}
\put(29.333333,41.357855){.}
\multiput(29.333333,0)(0,4){11}{\line(0,1){1}}
\put(27,-5){$\rho^{*}_1$}
\put(29.666667,41.355679){.}
\put(30.000000,41.338481){.}
\put(30.333333,41.306058){.}
\put(30.666667,41.258202){.}
\put(31.000000,41.194703){.}
\put(31.333333,41.115348){.}
\put(31.666667,41.019919){.}
\put(32.000000,40.908197){.}
\put(32.333333,40.779959){.}
\put(32.666667,40.634978){.}
\put(33.000000,40.473026){.}
\put(33.333333,40.293870){.}
\put(33.666667,40.097276){.}
\put(34.000000,39.883006){.}
\put(34.333333,39.650819){.}
\put(34.666667,39.400473){.}
\put(35.000000,39.131722){.}
\put(35.333333,38.844320){.}
\put(35.666667,38.538018){.}
\put(36.000000,38.212565){.}
\put(36.333333,37.867711){.}
\put(36.666667,37.503203){.}
\put(37.000000,37.118790){.}
\put(37.333333,36.714222){.}
\put(37.666667,36.289247){.}
\put(38.000000,35.843620){.}
\put(38.333333,35.377098){.}
\put(38.666667,34.889442){.}
\put(39.000000,34.380419){.}
\put(39.333333,33.849807){.}
\put(39.666667,33.297393){.}
\put(40.000000,32.722977){.}
\put(40.333333,32.126377){.}
\put(40.666667,31.507429){.}
\put(41.000000,30.865997){.}
\put(41.333333,30.201974){.}
\put(41.666667,29.515291){.}
\put(42.000000,28.805926){.}
\put(42.333333,28.073913){.}
\put(42.666667,27.319355){.}
\multiput(42,0)(0,4){7}{\line(0,1){1}}
\put(40,-5){$\hat \rho$}

\put(43.000000,26.542441){.}
\put(43.333333,25.743465){.}
\put(43.666667,24.922850){.}
\put(44.000000,24.081182){.}
\put(44.333333,23.219245){.}
\put(44.666667,22.338078){.}
\put(45.000000,21.439033){.}
\put(45.333333,20.523861){.}
\put(45.666667,19.594819){.}
\put(46.000000,18.654806){.}
\put(46.333333,17.707537){.}
\put(46.666667,16.757771){.}
\put(47.000000,15.811582){.}
\put(47.333333,14.876703){.}
\put(47.666667,13.962897){.}
\put(48.000000,13.082345){.}
\put(48.333333,12.249927){.}
\put(48.666667,11.483217){.}
\put(49.000000,10.801906){.}
\put(49.333333,10.226339){.}
\put(49.666667,9.775026){.}
\put(50.000000,9.461498){.}
\put(50.333333,9.291509){.}
\put(50.666667,9.261837){.}
\multiput(50.666667,0)(0,4){3}{\line(0,1){1}}
\put(47,-5){$\rho_{min}$}

\multiput(-0.5,9.261837)(4,0){24}{\line(1,0){1}}
\multiput(3.9,0)(0,4){3}{\line(0,1){1}}
\multiput(94.2,0)(0,4){3}{\line(0,1){1}}

\put(1.9,-5){$\tilde \rho_1$}
\put(92.2,-5){$\tilde \rho_2$}

\put(51.000000,9.361255){.}
\put(51.333333,9.573180){.}
\put(51.666667,9.878755){.}
\put(52.000000,10.259367){.}
\put(52.333333,10.698175){.}
\put(52.666667,11.180769){.}
\put(53.000000,11.695242){.}
\put(53.333333,12.231964){.}
\put(53.666667,12.783244){.}
\put(54.000000,13.342978){.}
\put(54.333333,13.906336){.}
\put(54.666667,14.469497){.}
\put(55.000000,15.029437){.}
\put(55.333333,15.583761){.}
\put(55.666667,16.130569){.}
\put(56.000000,16.668354){.}
\put(56.333333,17.195923){.}
\put(56.666667,17.712331){.}
\put(57.000000,18.216835){.}
\put(57.333333,18.708854){.}
\put(57.666667,19.187938){.}
\put(58.000000,19.653743){.}
\put(58.333333,20.106014){.}
\put(58.666667,20.544567){.}
\put(59.000000,20.969276){.}
\put(59.333333,21.380066){.}
\put(59.666667,21.776901){.}
\put(60.000000,22.159779){.}
\put(60.333333,22.528725){.}
\put(60.666667,22.883788){.}
\put(61.000000,23.225036){.}
\put(61.333333,23.552554){.}
\put(61.666667,23.866437){.}
\put(62.000000,24.166795){.}
\put(62.333333,24.453745){.}
\put(62.666667,24.727411){.}
\put(63.000000,24.987925){.}
\put(63.333333,25.235420){.}
\put(63.666667,25.470038){.}
\put(64.000000,25.691920){.}
\put(64.333333,25.901210){.}
\put(64.666667,26.098056){.}
\put(65.000000,26.282604){.}
\put(65.333333,26.455003){.}
\put(65.666667,26.615400){.}
\put(66.000000,26.763945){.}
\put(66.333333,26.900784){.}
\put(66.666667,27.026067){.}
\put(67.000000,27.139938){.}
\put(67.333333,27.242543){.}
\put(67.666667,27.334028){.}
\put(68.000000,27.414534){.}
\put(68.333333,27.484205){.}
\put(68.666667,27.543180){.}
\put(69.000000,27.591599){.}
\put(69.333333,27.629598){.}
\put(69.666667,27.657313){.}
\put(70.000000,27.674878){.}
\put(70.333333,27.682426){.}
\multiput(70.333333,0)(0,4){7}{\line(0,1){1}}
\put(67,-5){$\rho^{*}_2$}
\multiput(42.333333,27.682426)(4,0){8}{\line(1,0){1}}

\put(70.666667,27.680087){.}
\put(71.000000,27.667989){.}
\put(71.333333,27.646260){.}
\put(71.666667,27.615025){.}
\put(72.000000,27.574407){.}
\put(72.333333,27.524528){.}
\put(72.666667,27.465507){.}
\put(73.000000,27.397464){.}
\put(73.333333,27.320515){.}
\put(73.666667,27.234773){.}
\put(74.000000,27.140352){.}
\put(74.333333,27.037363){.}
\put(74.666667,26.925916){.}
\put(75.000000,26.806119){.}
\put(75.333333,26.678077){.}
\put(75.666667,26.541896){.}
\put(76.000000,26.397677){.}
\put(76.333333,26.245524){.}
\put(76.666667,26.085534){.}
\put(77.000000,25.917808){.}
\put(77.333333,25.742440){.}
\put(77.666667,25.559527){.}
\put(78.000000,25.369162){.}
\put(78.333333,25.171438){.}
\put(78.666667,24.966445){.}
\put(79.000000,24.754273){.}
\put(79.333333,24.535010){.}
\put(79.666667,24.308742){.}
\put(80.000000,24.075555){.}
\put(80.333333,23.835533){.}
\put(80.666667,23.588759){.}
\put(81.000000,23.335313){.}
\put(81.333333,23.075277){.}
\put(81.666667,22.808729){.}
\put(82.000000,22.535747){.}
\put(82.333333,22.256407){.}
\put(82.666667,21.970785){.}
\put(83.000000,21.678956){.}
\put(83.333333,21.380991){.}
\put(83.666667,21.076964){.}
\put(84.000000,20.766944){.}
\put(84.333333,20.451003){.}
\put(84.666667,20.129208){.}
\put(85.000000,19.801627){.}
\put(85.333333,19.468328){.}
\put(85.666667,19.129376){.}
\put(86.000000,18.784835){.}
\put(86.333333,18.434769){.}
\put(86.666667,18.079242){.}
\put(87.000000,17.718315){.}
\put(87.333333,17.352049){.}
\put(87.666667,16.980504){.}
\put(88.000000,16.603739){.}
\put(88.333333,16.221814){.}
\put(88.666667,15.834784){.}
\put(89.000000,15.442708){.}
\put(89.333333,15.045640){.}
\put(89.666667,14.643635){.}
\put(90.000000,14.236749){.}
\put(90.333333,13.825034){.}
\put(90.666667,13.408543){.}
\put(91.000000,12.987329){.}
\put(91.333333,12.561441){.}
\put(91.666667,12.130931){.}
\put(92.000000,11.695849){.}
\put(92.333333,11.256243){.}
\put(92.666667,10.812161){.}
\put(93.000000,10.363653){.}
\put(93.333333,9.910763){.}
\put(93.666667,9.453540){.}
\put(94.000000,8.992028){.}
\put(94.333333,8.526273){.}
\put(94.666667,8.056318){.}
\put(95.000000,7.582209){.}
\put(95.333333,7.103988){.}
\put(95.666667,6.621697){.}
\put(96.000000,6.135379){.}
\put(96.333333,5.645076){.}
\put(96.666667,5.150827){.}
\put(97.000000,4.652675){.}
\put(97.333333,4.150658){.}
\put(97.666667,3.644815){.}
\put(98.000000,3.135186){.}
\put(98.333333,2.621808){.}
\put(98.666667,2.104720){.}
\put(99.000000,1.583959){.}
\put(99.333333,1.059561){.}
\put(99.666667,0.531563){.}

\end{picture}
\end{center}
\caption{
Exact current-density relation of the TASEP with nearest-neighbor
interaction for $\epsilon=0.995,\ \delta=0.2$
(Eqs.(\ref{rate1})-(\ref{rate4})).}
\label{curr}
\end{figure}

\bigskip
\vspace{3truecm}

\input{euro_fig2.tex}

\bigskip

\newpage
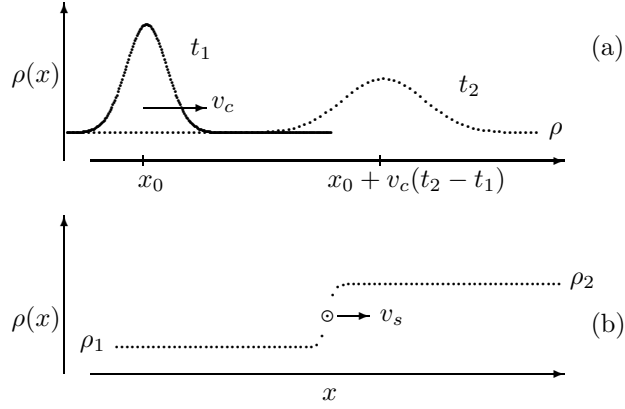
\begin{figure}
\setlength{\unitlength}{0.7mm}
\begin{center}
\begin{picture}(100,30)(0,0)

\put(100,25){(a)}

\put(0,5){\vector(0,1){30}}
\put(-10,20){$\rho(x)$}
\put(92,10){$\rho$}

\put(5,5){\vector(1,0){90}}
\put(14,0){$x_0$}
\put(15,4){\line(0,1){2}}
\put(50,0){$x_0 + v_c(t_2 - t_1)$}
\put(60,4){\line(0,1){2}}

\put(24,25){$t_1$}
\put(75,18){$t_2$}
\put(15,15){\vector(1,0){12}}
\put(28,15){$v_c$}

\put(0,10.){.}
\put(0.2,10.){.}
\put(0.4,10.){.}
\put(0.6,10.){.}
\put(0.8,10.){.}
\put(1.,10.){.}
\put(1.2,10.){.}
\put(1.4,10.){.}
\put(1.6,10.1){.}
\put(1.8,10.1){.}
\put(2.,10.1){.}
\put(2.2,10.1){.}
\put(2.4,10.1){.}
\put(2.6,10.1){.}
\put(2.8,10.1){.}
\put(3.,10.2){.}
\put(3.2,10.2){.}
\put(3.4,10.2){.}
\put(3.6,10.3){.}
\put(3.8,10.3){.}
\put(4.,10.4){.}
\put(4.2,10.4){.}
\put(4.4,10.5){.}
\put(4.6,10.6){.}
\put(4.8,10.6){.}
\put(5.,10.7){.}
\put(5.2,10.8){.}
\put(5.4,11.){.}
\put(5.6,11.1){.}
\put(5.8,11.2){.}
\put(6.,11.4){.}
\put(6.2,11.6){.}
\put(6.4,11.8){.}
\put(6.6,12.){.}
\put(6.8,12.2){.}
\put(7.,12.4){.}
\put(7.2,12.7){.}
\put(7.4,13.){.}
\put(7.6,13.3){.}
\put(7.8,13.7){.}
\put(8.,14.){.}
\put(8.2,14.4){.}
\put(8.4,14.8){.}
\put(8.6,15.3){.}
\put(8.8,15.7){.}
\put(9.,16.2){.}
\put(9.2,16.7){.}
\put(9.4,17.2){.}
\put(9.6,17.8){.}
\put(9.8,18.4){.}
\put(10.,19.){.}
\put(10.2,19.6){.}
\put(10.4,20.2){.}
\put(10.6,20.8){.}
\put(10.8,21.4){.}
\put(11.,22.1){.}
\put(11.2,22.7){.}
\put(11.4,23.4){.}
\put(11.6,24.){.}
\put(11.8,24.6){.}
\put(12.,25.3){.}
\put(12.2,25.9){.}
\put(12.4,26.4){.}
\put(12.6,27.){.}
\put(12.8,27.5){.}
\put(13.,28.){.}
\put(13.2,28.5){.}
\put(13.4,28.9){.}
\put(13.6,29.3){.}
\put(13.8,29.6){.}
\put(14.,29.9){.}
\put(14.2,30.2){.}
\put(14.4,30.4){.}
\put(14.6,30.5){.}
\put(14.8,30.6){.}
\put(15.,30.6){.}
\put(15.2,30.6){.}
\put(15.4,30.5){.}
\put(15.6,30.4){.}
\put(15.8,30.2){.}
\put(16.,29.9){.}
\put(16.2,29.6){.}
\put(16.4,29.3){.}
\put(16.6,28.9){.}
\put(16.8,28.5){.}
\put(17.,28.){.}
\put(17.2,27.5){.}
\put(17.4,27.){.}
\put(17.6,26.4){.}
\put(17.8,25.9){.}
\put(18.,25.3){.}
\put(18.2,24.6){.}
\put(18.4,24.){.}
\put(18.6,23.4){.}
\put(18.8,22.7){.}
\put(19.,22.1){.}
\put(19.2,21.4){.}
\put(19.4,20.8){.}
\put(19.6,20.2){.}
\put(19.8,19.6){.}
\put(20.,19.){.}
\put(20.2,18.4){.}
\put(20.4,17.8){.}
\put(20.6,17.2){.}
\put(20.8,16.7){.}
\put(21.,16.2){.}
\put(21.2,15.7){.}
\put(21.4,15.3){.}
\put(21.6,14.8){.}
\put(21.8,14.4){.}
\put(22.,14.){.}
\put(22.2,13.7){.}
\put(22.4,13.3){.}
\put(22.6,13.){.}
\put(22.8,12.7){.}
\put(23.,12.4){.}
\put(23.2,12.2){.}
\put(23.4,12.){.}
\put(23.6,11.8){.}
\put(23.8,11.6){.}
\put(24.,11.4){.}
\put(24.2,11.2){.}
\put(24.4,11.1){.}
\put(24.6,11.){.}
\put(24.8,10.8){.}
\put(25.,10.7){.}
\put(25.2,10.6){.}
\put(25.4,10.6){.}
\put(25.6,10.5){.}
\put(25.8,10.4){.}
\put(26.,10.4){.}
\put(26.2,10.3){.}
\put(26.4,10.3){.}
\put(26.6,10.2){.}
\put(26.8,10.2){.}
\put(27.,10.2){.}
\put(27.2,10.1){.}
\put(27.4,10.1){.}
\put(27.6,10.1){.}
\put(27.8,10.1){.}
\put(28.,10.1){.}
\put(28.2,10.1){.}
\put(28.4,10.1){.}
\put(28.6,10.){.}
\put(28.8,10.){.}
\put(29.,10.){.}
\put(29.2,10.){.}
\put(29.4,10.){.}
\put(29.6,10.){.}
\put(29.8,10.){.}
\put(30.,10.){.}
\put(30.2,10.){.}
\put(30.4,10.){.}
\put(30.6,10.){.}
\put(30.8,10.){.}
\put(31.,10.){.}
\put(31.2,10.){.}
\put(31.4,10.){.}
\put(31.6,10.){.}
\put(31.8,10.){.}
\put(32.,10.){.}
\put(32.2,10.){.}
\put(32.4,10.){.}
\put(32.6,10.){.}
\put(32.8,10.){.}
\put(33.,10.){.}
\put(33.2,10.){.}
\put(33.4,10.){.}
\put(33.6,10.){.}
\put(33.8,10.){.}
\put(34.,10.){.}
\put(34.2,10.){.}
\put(34.4,10.){.}
\put(34.6,10.){.}
\put(34.8,10.){.}
\put(35.,10.){.}
\put(35.2,10.){.}
\put(35.4,10.){.}
\put(35.6,10.){.}
\put(35.8,10.){.}
\put(36.,10.){.}
\put(36.2,10.){.}
\put(36.4,10.){.}
\put(36.6,10.){.}
\put(36.8,10.){.}
\put(37.,10.){.}
\put(37.2,10.){.}
\put(37.4,10.){.}
\put(37.6,10.){.}
\put(37.8,10.){.}
\put(38.,10.){.}
\put(38.2,10.){.}
\put(38.4,10.){.}
\put(38.6,10.){.}
\put(38.8,10.){.}
\put(39.,10.){.}
\put(39.2,10.){.}
\put(39.4,10.){.}
\put(39.6,10.){.}
\put(39.8,10.){.}
\put(40.,10.){.}
\put(40.2,10.){.}
\put(40.4,10.){.}
\put(40.6,10.){.}
\put(40.8,10.){.}
\put(41.,10.){.}
\put(41.2,10.){.}
\put(41.4,10.){.}
\put(41.6,10.){.}
\put(41.8,10.){.}
\put(42.,10.){.}
\put(42.2,10.){.}
\put(42.4,10.){.}
\put(42.6,10.){.}
\put(42.8,10.){.}
\put(43.,10.){.}
\put(43.2,10.){.}
\put(43.4,10.){.}
\put(43.6,10.){.}
\put(43.8,10.){.}
\put(44.,10.){.}
\put(44.2,10.){.}
\put(44.4,10.){.}
\put(44.6,10.){.}
\put(44.8,10.){.}
\put(45.,10.){.}
\put(45.2,10.){.}
\put(45.4,10.){.}
\put(45.6,10.){.}
\put(45.8,10.){.}
\put(46.,10.){.}
\put(46.2,10.){.}
\put(46.4,10.){.}
\put(46.6,10.){.}
\put(46.8,10.){.}
\put(47.,10.){.}
\put(47.2,10.){.}
\put(47.4,10.){.}
\put(47.6,10.){.}
\put(47.8,10.){.}
\put(48.,10.){.}
\put(48.2,10.){.}
\put(48.4,10.){.}
\put(48.6,10.){.}
\put(48.8,10.){.}
\put(49.,10.){.}
\put(49.2,10.){.}
\put(49.4,10.){.}
\put(49.6,10.){.}
\put(49.8,10.){.}
\put(50.,10.){.}
\put(0,10.){.}
\put(1.,10.){.}
\put(2.,10.){.}
\put(3.,10.){.}
\put(4.,10.){.}
\put(5.,10.){.}
\put(6.,10.){.}
\put(7.,10.){.}
\put(8.,10.){.}
\put(9.,10.){.}
\put(10.,10.){.}
\put(11.,10.){.}
\put(12.,10.){.}
\put(13.,10.){.}
\put(14.,10.){.}
\put(15.,10.){.}
\put(16.,10.){.}
\put(17.,10.){.}
\put(18.,10.){.}
\put(19.,10.){.}
\put(20.,10.){.}
\put(21.,10.){.}
\put(22.,10.){.}
\put(23.,10.){.}
\put(24.,10.){.}
\put(25.,10.){.}
\put(26.,10.){.}
\put(27.,10.){.}
\put(28.,10.){.}
\put(29.,10.){.}
\put(30.,10.){.}
\put(31.,10.){.}
\put(32.,10.){.}
\put(33.,10.){.}
\put(34.,10.){.}
\put(35.,10.1){.}
\put(36.,10.1){.}
\put(37.,10.1){.}
\put(38.,10.2){.}
\put(39.,10.3){.}
\put(40.,10.4){.}
\put(41.,10.5){.}
\put(42.,10.7){.}
\put(43.,10.9){.}
\put(44.,11.2){.}
\put(45.,11.6){.}
\put(46.,12.){.}
\put(47.,12.5){.}
\put(48.,13.1){.}
\put(49.,13.8){.}
\put(50.,14.5){.}
\put(51.,15.2){.}
\put(52.,16.){.}
\put(53.,16.8){.}
\put(54.,17.6){.}
\put(55.,18.4){.}
\put(56.,19.){.}
\put(57.,19.6){.}
\put(58.,20.){.}
\put(59.,20.2){.}
\put(60.,20.3){.}
\put(61.,20.2){.}
\put(62.,20.){.}
\put(63.,19.6){.}
\put(64.,19.){.}
\put(65.,18.4){.}
\put(66.,17.6){.}
\put(67.,16.8){.}
\put(68.,16.){.}
\put(69.,15.2){.}
\put(70.,14.5){.}
\put(71.,13.8){.}
\put(72.,13.1){.}
\put(73.,12.5){.}
\put(74.,12.){.}
\put(75.,11.6){.}
\put(76.,11.2){.}
\put(77.,10.9){.}
\put(78.,10.7){.}
\put(79.,10.5){.}
\put(80.,10.4){.}
\put(81.,10.3){.}
\put(82.,10.2){.}
\put(83.,10.1){.}
\put(84.,10.1){.}
\put(85.,10.1){.}
\put(86.,10.){.}
\put(87.,10.){.}
\put(88.,10.){.}
\put(89.,10.){.}

\end{picture}

\vspace*{7mm}
\begin{picture}(120,30)

\put(110,13){(b)}

\put(60,16){\circle{2.}}
\put(62,16){\vector(1,0){6.}}
\put(70,15){$v_s$}

\put(0,14){$\rho(x)$}
\put(10,5){\vector(0,1){30}}
\put(59,0){$x$}
\put(15,5){\vector(1,0){90}}
\put(13,10){$\rho_{1}$}
\put(106,22){$\rho_{2}$}

\put(20.,10.){\circle*{.7}}
\put(21.,10.){\circle*{.7}}
\put(22.,10.){\circle*{.7}}
\put(23.,10.){\circle*{.7}}
\put(24.,10.){\circle*{.7}}
\put(25.,10.){\circle*{.7}}
\put(26.,10.){\circle*{.7}}
\put(27.,10.){\circle*{.7}}
\put(28.,10.){\circle*{.7}}
\put(29.,10.){\circle*{.7}}
\put(30.,10.){\circle*{.7}}
\put(31.,10.){\circle*{.7}}
\put(32.,10.){\circle*{.7}}
\put(33.,10.){\circle*{.7}}
\put(34.,10.){\circle*{.7}}
\put(35.,10.){\circle*{.7}}
\put(36.,10.){\circle*{.7}}
\put(37.,10.){\circle*{.7}}
\put(38.,10.){\circle*{.7}}
\put(39.,10.){\circle*{.7}}
\put(40.,10.){\circle*{.7}}
\put(41.,10.){\circle*{.7}}
\put(42.,10.){\circle*{.7}}
\put(43.,10.){\circle*{.7}}
\put(44.,10.){\circle*{.7}}
\put(45.,10.){\circle*{.7}}
\put(46.,10.){\circle*{.7}}
\put(47.,10.){\circle*{.7}}
\put(48.,10.){\circle*{.7}}
\put(49.,10.){\circle*{.7}}
\put(50.,10.){\circle*{.7}}
\put(51.,10.){\circle*{.7}}
\put(52.,10.){\circle*{.7}}
\put(53.,10.){\circle*{.7}}
\put(54.,10.){\circle*{.7}}
\put(55.,10.){\circle*{.7}}
\put(56.,10.){\circle*{.7}}
\put(57.,10.1){\circle*{.7}}
\put(58.,10.5){\circle*{.7}}
\put(59.,12.1){\circle*{.7}}
\put(60.,16.){\circle*{.7}}
\put(61.,19.9){\circle*{.7}}
\put(62.,21.5){\circle*{.7}}
\put(63.,21.9){\circle*{.7}}
\put(64.,22.){\circle*{.7}}
\put(65.,22.){\circle*{.7}}
\put(66.,22.){\circle*{.7}}
\put(67.,22.){\circle*{.7}}
\put(68.,22.){\circle*{.7}}
\put(69.,22.){\circle*{.7}}
\put(70.,22.){\circle*{.7}}
\put(71.,22.){\circle*{.7}}
\put(72.,22.){\circle*{.7}}
\put(73.,22.){\circle*{.7}}
\put(74.,22.){\circle*{.7}}
\put(75.,22.){\circle*{.7}}
\put(76.,22.){\circle*{.7}}
\put(77.,22.){\circle*{.7}}
\put(78.,22.){\circle*{.7}}
\put(79.,22.){\circle*{.7}}
\put(80.,22.){\circle*{.7}}
\put(81.,22.){\circle*{.7}}
\put(82.,22.){\circle*{.7}}
\put(83.,22.){\circle*{.7}}
\put(84.,22.){\circle*{.7}}
\put(85.,22.){\circle*{.7}}
\put(86.,22.){\circle*{.7}}
\put(87.,22.){\circle*{.7}}
\put(88.,22.){\circle*{.7}}
\put(89.,22.){\circle*{.7}}
\put(90.,22.){\circle*{.7}}
\put(91.,22.){\circle*{.7}}
\put(92.,22.){\circle*{.7}}
\put(93.,22.){\circle*{.7}}
\put(94.,22.){\circle*{.7}}
\put(95.,22.){\circle*{.7}}
\put(96.,22.){\circle*{.7}}
\put(97.,22.){\circle*{.7}}
\put(98.,22.){\circle*{.7}}
\put(99.,22.){\circle*{.7}}
\put(100.,22.){\circle*{.7}}
\put(101.,22.){\circle*{.7}}
\put(102.,22.){\circle*{.7}}
\put(103.,22.){\circle*{.7}}
\put(104.,22.){\circle*{.7}}

\end{picture}
\end{center}
\caption{\small (a)
Diffusive spreading of a density perturbation in the stationary state
at two times $t_2 > t_1$. The collective velocity describes the motion
of the center of mass of the perturbation. (b) Motion of a shock. To the left
of the domain wall particles are distributed
homogeneously with an average density $\rho_{1}$. To the
right of the domain wall the background density is $\rho_{2}$}
\label{v}
\end{figure}

\vspace{2.5truecm}

\begin{figure}
\setlength{\unitlength}{0.6mm}
\begin{center}
%\begin{picture}(100,40)(0,-10)
\begin{picture}(100,30)(0,-7)

\put(-5,5){\vector(0,1){20}}
\put(5,-5){\vector(1,0){90}}

\put(50,-8){$x$}
\put(-10,15){$\rho$}

%\multiput(100,7.5)(0,7.5){3}{\line(-1,0){1}}
\put(5,0){\line(1,0){25}}
%\multiput(30,0)(5.6568542,5.6568542){6}{\line(0,1){4}}
%\multiput(30,4)(5.6568542,5.6568542){5}{\line(1,0){4}}
\multiput(30,0)(4,4){6}{\line(0,1){4}}
\multiput(30,4)(4,4){5}{\line(1,0){4}}
\put(50,24){\line(1,0){30}}

\multiput(29,2)(4,4){2}{\vector(-1,0){5}}
\multiput(39,10)(4,4){4}{\vector(1,0){5}}

\put(0,-4){$\rho^-$}
\put(82,22){$\rho^+$}

\multiput(10,8)(5,0){11}{\line(1,0){1}}
\put(62,6){$\rho_{min}$}

\end{picture}

\end{center}
\caption{Schematical drawing of the decomposition of a large shock into
small shocks and their velocities, leading to branching and coalescence.
Here $\rho^{*}_1< \rho^- < \rho_{min} < \rho^+<\rho^{*}_2$.
}
\label{schema}
\end{figure}
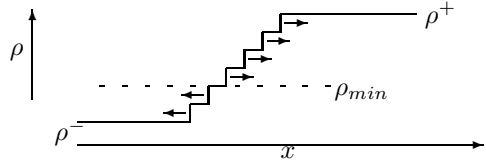

\vspace{2.5truecm}

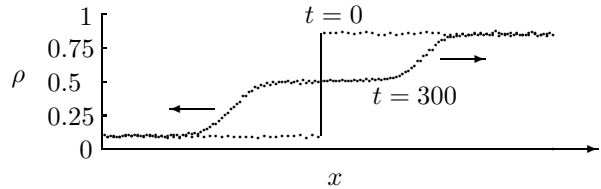
\begin{figure}
\setlength{\unitlength}{0.6mm}
\begin{center}
%\begin{picture}(100,40)(0,-10)
\begin{picture}(100,35)(0,-7)

\put(0,0){\line(0,1){30}}
%\put(100,0){\line(0,1){30}}
\put(0,0){\vector(1,0){110}}
%\put(0,30){\line(1,0){100}}

\put(50,-8){$x$}
\put(-20,15){$\rho$}

\multiput(0,7.5)(0,7.5){4}{\line(1,0){1}}
\put(-5,-2){$0$}
\put(-12,5.5){$0.25$}
\put(-12,13){$0.5$}
\put(-12,20.5){$0.75$}
\put(-5,27.5){$1$}

%\multiput(100,7.5)(0,7.5){3}{\line(-1,0){1}}

\put(75,20){\vector(1,0){10}}
\put(25,9){\vector(-1,0){10}}
\put(45,28){$t=0$}
\put(60,10){$t=300$}

\put(1.33333,2.79){\circle*{0.5}}
\put(2.66667,2.72001){\circle*{0.5}}
\put(4,2.78001){\circle*{0.5}}
\put(5.33333,3.03){\circle*{0.5}}
\put(6.66667,2.94999){\circle*{0.5}}
\put(8,2.87001){\circle*{0.5}}
\put(9.33333,2.93001){\circle*{0.5}}
\put(10.6667,2.93001){\circle*{0.5}}
\put(12,2.63001){\circle*{0.5}}
\put(13.3333,2.91){\circle*{0.5}}
\put(14.6667,2.61){\circle*{0.5}}
\put(16,2.73){\circle*{0.5}}
\put(17.3333,2.60001){\circle*{0.5}}
\put(18.6667,2.85){\circle*{0.5}}
\put(20,2.81001){\circle*{0.5}}
\put(21.3333,2.85){\circle*{0.5}}
\put(22.6667,2.7){\circle*{0.5}}
\put(24,2.82999){\circle*{0.5}}
\put(25.3333,2.90001){\circle*{0.5}}
\put(26.6667,2.64){\circle*{0.5}}
\put(28,2.93001){\circle*{0.5}}
\put(29.3333,2.72001){\circle*{0.5}}
\put(30.6667,2.7){\circle*{0.5}}
\put(32,2.79){\circle*{0.5}}
\put(33.3333,2.81001){\circle*{0.5}}
\put(34.6667,2.58){\circle*{0.5}}
\put(36,3.03999){\circle*{0.5}}
\put(37.3333,2.69001){\circle*{0.5}}
\put(38.6667,2.76999){\circle*{0.5}}
\put(40,2.85){\circle*{0.5}}
\put(41.3333,2.88){\circle*{0.5}}
\put(42.6667,2.84001){\circle*{0.5}}
\put(44,3.11001){\circle*{0.5}}
\put(45.3333,2.49){\circle*{0.5}}
\put(46.6667,3.14001){\circle*{0.5}}
\put(48,2.76){\circle*{0.5}}

\put(48.6,3){\line(0,1){22.76}}

\put(49.3333,25.76){\circle*{0.5}}
\put(50.6667,25.71){\circle*{0.5}}
\put(52,25.78){\circle*{0.5}}
\put(53.3333,26.02){\circle*{0.5}}
\put(54.6667,25.85){\circle*{0.5}}
\put(56,25.52){\circle*{0.5}}
\put(57.3333,25.87){\circle*{0.5}}
\put(58.6667,26.05){\circle*{0.5}}
\put(60,25.76){\circle*{0.5}}
\put(61.3333,25.32){\circle*{0.5}}
\put(62.6667,25.47){\circle*{0.5}}
\put(64,25.69){\circle*{0.5}}
\put(65.3333,25.44){\circle*{0.5}}
\put(66.6667,25.63){\circle*{0.5}}
\put(68,25.75){\circle*{0.5}}
\put(69.3333,25.48){\circle*{0.5}}
\put(70.6667,25.78){\circle*{0.5}}
\put(72,25.99){\circle*{0.5}}
\put(73.3333,25.79){\circle*{0.5}}
\put(74.6667,25.83){\circle*{0.5}}
\put(76,25.44){\circle*{0.5}}
\put(77.3333,25.39){\circle*{0.5}}
\put(78.6667,25.68){\circle*{0.5}}
\put(80,25.61){\circle*{0.5}}
\put(81.3333,25.35){\circle*{0.5}}
\put(82.6667,25.78){\circle*{0.5}}
\put(84,25.51){\circle*{0.5}}
\put(85.3333,25.67){\circle*{0.5}}
\put(86.6667,25.74){\circle*{0.5}}
\put(88,25.78){\circle*{0.5}}
\put(89.3333,25.62){\circle*{0.5}}
\put(90.6667,25.81){\circle*{0.5}}
\put(92,25.8){\circle*{0.5}}
\put(93.3333,25.27){\circle*{0.5}}
\put(94.6667,25.72){\circle*{0.5}}
\put(96,25.7){\circle*{0.5}}
\put(97.3333,25.44){\circle*{0.5}}
\put(98.6667,25.65){\circle*{0.5}}
\put(100,0){\circle*{0.5}}

\put(0.666667,3.24999){\circle*{0.7}}
\put(1.33333,2.73){\circle*{0.7}}
\put(2,2.96001){\circle*{0.7}}
\put(2.66667,3.00999){\circle*{0.7}}
\put(3.33333,2.76){\circle*{0.7}}
\put(4,2.82){\circle*{0.7}}
\put(4.66667,2.97){\circle*{0.7}}
\put(5.33333,3.09999){\circle*{0.7}}
\put(6,3.06){\circle*{0.7}}
\put(6.66667,2.91){\circle*{0.7}}
\put(7.33333,3.15999){\circle*{0.7}}
\put(8,3.08001){\circle*{0.7}}
\put(8.66667,3.02001){\circle*{0.7}}
\put(9.33333,2.7){\circle*{0.7}}
\put(10,3.30999){\circle*{0.7}}
\put(10.6667,3.3){\circle*{0.7}}
\put(11.3333,3.03999){\circle*{0.7}}
\put(12,3.27999){\circle*{0.7}}
\put(12.6667,2.73){\circle*{0.7}}
\put(13.3333,2.94){\circle*{0.7}}
\put(14,2.79999){\circle*{0.7}}
\put(14.6667,2.88){\circle*{0.7}}
\put(15.3333,3.18999){\circle*{0.7}}
\put(16,3.20001){\circle*{0.7}}
\put(16.6667,3.09){\circle*{0.7}}
\put(17.3333,2.7){\circle*{0.7}}
\put(18,3.3){\circle*{0.7}}
\put(18.6667,3.3){\circle*{0.7}}
\put(19.3333,3.72999){\circle*{0.7}}
\put(20,3.41001){\circle*{0.7}}
\put(20.6667,3.72999){\circle*{0.7}}
\put(21.3333,3.48){\circle*{0.7}}
\put(22,3.83001){\circle*{0.7}}
\put(22.6667,4.32999){\circle*{0.7}}
\put(23.3333,4.44){\circle*{0.7}}
\put(24,5.04999){\circle*{0.7}}
\put(24.6667,5.52999){\circle*{0.7}}
\put(25.3333,5.61){\circle*{0.7}}
\put(26,6.32001){\circle*{0.7}}
\put(26.6667,6.9){\circle*{0.7}}
\put(27.3333,7.59){\circle*{0.7}}
\put(28,8.1){\circle*{0.7}}
\put(28.6667,8.99001){\circle*{0.7}}
\put(29.3333,9.27999){\circle*{0.7}}
\put(30,10.12){\circle*{0.7}}
\put(30.6667,10.43){\circle*{0.7}}
\put(31.3333,11.92){\circle*{0.7}}
\put(32,11.89){\circle*{0.7}}
\put(32.6667,12.7){\circle*{0.7}}
\put(33.3333,12.91){\circle*{0.7}}
\put(34,13.76){\circle*{0.7}}
\put(34.6667,13.55){\circle*{0.7}}
\put(35.3333,14.3){\circle*{0.7}}
\put(36,14.09){\circle*{0.7}}
\put(36.6667,14.67){\circle*{0.7}}
\put(37.3333,14.29){\circle*{0.7}}
\put(38,14.89){\circle*{0.7}}
\put(38.6667,14.52){\circle*{0.7}}
\put(39.3333,14.97){\circle*{0.7}}
\put(40,14.91){\circle*{0.7}}
\put(40.6667,14.83){\circle*{0.7}}
\put(41.3333,15.04){\circle*{0.7}}
\put(42,14.82){\circle*{0.7}}
\put(42.6667,14.92){\circle*{0.7}}
\put(43.3333,15.23){\circle*{0.7}}
\put(44,14.55){\circle*{0.7}}
\put(44.6667,15.32){\circle*{0.7}}
\put(45.3333,14.75){\circle*{0.7}}
\put(46,15.22){\circle*{0.7}}
\put(46.6667,14.78){\circle*{0.7}}
\put(47.3333,15.11){\circle*{0.7}}
\put(48,14.93){\circle*{0.7}}
\put(48.6667,15.17){\circle*{0.7}}
\put(49.3333,15.15){\circle*{0.7}}
\put(50,15.11){\circle*{0.7}}
\put(50.6667,15.27){\circle*{0.7}}
\put(51.3333,15.06){\circle*{0.7}}
\put(52,15.22){\circle*{0.7}}
\put(52.6667,15.05){\circle*{0.7}}
\put(53.3333,15.33){\circle*{0.7}}
\put(54,15.15){\circle*{0.7}}
\put(54.6667,15.19){\circle*{0.7}}
\put(55.3333,15.15){\circle*{0.7}}
\put(56,15.41){\circle*{0.7}}
\put(56.6667,15.12){\circle*{0.7}}
\put(57.3333,15.37){\circle*{0.7}}
\put(58,15.09){\circle*{0.7}}
\put(58.6667,15.51){\circle*{0.7}}
\put(59.3333,15.19){\circle*{0.7}}
\put(60,15.51){\circle*{0.7}}
\put(60.6667,15.39){\circle*{0.7}}
\put(61.3333,15.58){\circle*{0.7}}
\put(62,15.28){\circle*{0.7}}
\put(62.6667,15.49){\circle*{0.7}}
\put(63.3333,15.48){\circle*{0.7}}
\put(64,15.77){\circle*{0.7}}
\put(64.6667,15.88){\circle*{0.7}}
\put(65.3333,16){\circle*{0.7}}
\put(66,16.29){\circle*{0.7}}
\put(66.6667,16.59){\circle*{0.7}}
\put(67.3333,17.2){\circle*{0.7}}
\put(68,17.38){\circle*{0.7}}
\put(68.6667,18.07){\circle*{0.7}}
\put(69.3333,18.55){\circle*{0.7}}
\put(70,19.55){\circle*{0.7}}
\put(70.6667,19.49){\circle*{0.7}}
\put(71.3333,20.72){\circle*{0.7}}
\put(72,20.92){\circle*{0.7}}
\put(72.6667,22.04){\circle*{0.7}}
\put(73.3333,22.06){\circle*{0.7}}
\put(74,23.06){\circle*{0.7}}
\put(74.6667,23.84){\circle*{0.7}}
\put(75.3333,24.28){\circle*{0.7}}
\put(76,24.28){\circle*{0.7}}
\put(76.6667,24.89){\circle*{0.7}}
\put(77.3333,24.94){\circle*{0.7}}
\put(78,24.78){\circle*{0.7}}
\put(78.6667,25.13){\circle*{0.7}}
\put(79.3333,25.25){\circle*{0.7}}
\put(80,25.61){\circle*{0.7}}
\put(80.6667,25.21){\circle*{0.7}}
\put(81.3333,25.46){\circle*{0.7}}
\put(82,25.42){\circle*{0.7}}
\put(82.6667,25.68){\circle*{0.7}}
\put(83.3333,25.46){\circle*{0.7}}
\put(84,25.73){\circle*{0.7}}
\put(84.6667,25.38){\circle*{0.7}}
\put(85.3333,25.48){\circle*{0.7}}
\put(86,25.3){\circle*{0.7}}
\put(86.6667,25.37){\circle*{0.7}}
\put(87.3333,26.15){\circle*{0.7}}
\put(88,25.07){\circle*{0.7}}
\put(88.6667,25.63){\circle*{0.7}}
\put(89.3333,25.38){\circle*{0.7}}
\put(90,25.65){\circle*{0.7}}
\put(90.6667,25.24){\circle*{0.7}}
\put(91.3333,25.45){\circle*{0.7}}
\put(92,25.59){\circle*{0.7}}
\put(92.6667,25.43){\circle*{0.7}}
\put(93.3333,25.32){\circle*{0.7}}
\put(94,25.46){\circle*{0.7}}
\put(94.6667,25.38){\circle*{0.7}}
\put(95.3333,26.05){\circle*{0.7}}
\put(96,25.6){\circle*{0.7}}
\put(96.6667,25.2){\circle*{0.7}}
\put(97.3333,25.37){\circle*{0.7}}
\put(98,25.59){\circle*{0.7}}
\put(98.6667,25.17){\circle*{0.7}}
\put(99.3333,25.84){\circle*{0.7}}
\put(100,25.33){\circle*{0.7}}

\end{picture}

\end{center}
\caption{ Snapshots of a particle density distribution at the initial
moment of time and after 300 Monte-Carlo steps, showing expansion
of the minimal current phase. Simulated is the system of 150 sites,
with particles initially distributed with average density
$\rho^{-}=0.1$ ($\rho^{+}=0.85$) on the left (on the right).
3000 different histories are averaged over.
}
\label{shock}
\end{figure}


\begin{references}
\bibitem{N1} We shall
consider here systems with a single local conservation law, i.e. which have a
family of bulk stationary states characterized by the particle density $\rho$.
Without loss of generality we assume a bias to the
right throughout this paper, i.e. a positive stationary particle current.
\bibitem{TASEP} The TASEP is a
lattice gas where each particle moves to the right with unit
rate (and exponential waiting-time distribution) if the neighboring site
is empty. Otherwise the move is rejected. At site 1 particles are injected
with rate $\alpha=\rho^-$ and at site $L$ particles are removed with $\beta
=1-\rho^+$. The bulk stationary states are uncorrelated and hence the
current-density relation is given by $j = \rho(1-\rho)$, with a maximum
$j^\ast=1/4$ at $\rho^\ast = 1/2$ \cite{Schm95}.
\bibitem{maxcoex} We consider the situation where the first maximum in
the current-density relation at density $\rho_1^{*}$ is higher than the
second maximum at $\rho_2^{*}$. The opposite case can be treated analogously.
In the case of degenerate maxima the two
maximal current phases have additional transitions to a maximal-current
coexistence phase in the $\rho^- > \rho^{*}_2$, $\rho^+ < \rho^{*}_1$
section of the phase diagram.
The spatial structure of the steady state
can be described as a superposition of two regimes with bulk densities
$\rho^{*}_1$ and $\rho^{*}_2$ resp., separated by a slowly fluctuating
domain wall which performs an unbiased random walk.
\end{references}
\end{document}